\documentstyle[preprint,aps]{revtex}
\begin{document}
\draft
\preprint{{\bf MITH 99/1}\\ \\ }
\title{ 
A FEW SIMPLE OBSERVATIONS ON PION-CONDENSATION IN NUCLEI
}
\author{{\bf R. Alzetta}}
\address{
Dipartimento di Fisica, Universit\`a della Calabria\\
INFN, Gruppo Collegato di Cosenza, I-87036 Rende (CS), Italy}
\author{{\bf G. Liberti}}
\address{
INFM, Unit\`a di Ricerca di Cosenza, I-87036 Rende (CS), Italy}
\author{{\bf G. Preparata}}
\address{
Dipartimento di Fisica, Universit\`a di Milano\\
INFN, Sezione di Milano, Via Celoria 16, I-20133 Milano, Italy}
\maketitle
\begin{abstract}
We present a few simple observations on the occurrence of 
$\pi$-condensation in Nuclei, aimed at clarifying the nature of the 
$\pi$-condensation implied by the coherent nuclear $\pi$-N-$\Delta$ 
interaction, proposed in 1990 to explain the puzzling emergence of 
the Shell-Model. We show that such condensation is totally unrelated 
to the one proposed by A. B. Migdal at the beginning of '70, which 
can easily be shown not to occur at the normal nucleon density 
$\rho_{N}\simeq$ 0.17 fm${}^{-3}$.
\end{abstract}
\bigskip
\vfill\eject

In 1990 a new approach to the dynamics of the Nucleus and of the Nuclear 
matter was proposed, based on an analogy between the \textit{coherent} 
$\pi$-Nucleon QCD interaction in nuclear matter and the coherent QED 
interaction in ordinary condensed matter \cite{Preparata1,Preparata2}.\par
The rapidly growing research program aimed at 
elucidating the r\^{o}le of the coherent electrodynamical 
interactions among the constituents (atoms and molecules) of ordinary 
condensed matter, lent itself in a surprisingly natural way to a far 
reaching generalization to nuclear matter, that could finally 
clarify several points of Nuclear Physics that had remained mysterious at 
least to the natural philosopher, if not to the expert of the field. 
The mysteries we are referring to can all be essentially encapsulated 
in the following question: why is the Shell-Model (SM) such a good 
(approximate) description of the structure of the Nucleus ? A question 
that has puzzled the more thoughtful students of Nuclear Physics, 
since its proposal by Mayer and Jensen almost 50 years ago 
\cite{Maier}.\par
Let's analyze the origin and motivations of the puzzle which the SM 
poses to our physical intuition. Since the seminal 
ideas of Yukawa \cite{Yukawa} nobody has ever put in doubt the notion 
that the nucleons of the Nucleus are held together by a 
``nucleostatic force'', the Yukawa interaction, arising from the 
virtual exchange of $\pi$-mesons between pairs of nucleons. The 
finiteness of the $\pi$-mass, as well known, implies the exponential 
decay of such force as $e^{-m_{\pi}r}$, giving it the rather short 
range $R_{1\pi}={1\over{m_{\pi}}}\simeq$ 1.4 fm. It is also well known 
that the basic $\pi$-exchange interaction has an important 
spin-isospin structure, which leads to repulsion instead of attraction 
in well defined spin-isospin channels. The same can be said of all other 
one-boson-exchanges which have, however, much smaller ranges 
\cite{Machleidt}. Thus the only universal dynamical mechanism of attraction 
between nucleons, irrespective of their spin and isospin, responsible for 
the existence of highly complex Nuclei, has been identified in the 
2$\pi$-exchange, involving the virtual transition to the 
$\Delta$(1232) as well. The range of this kind of nuclear Van der 
Waals forces is thus $R_{2\pi}={1\over{2 m_{\pi}}}\simeq$ 0.7 fm, a 
remarkably small distance approximately equal to the radius of the 
nucleons.\par
Let us now take an assembly of nucleons and squeeze them in a volume 
so small that their average mutual distance is comparable with 
$R_{2\pi}$\footnote{As a matter of fact for the actual average nucleon 
density $\rho_{N}\simeq$ 0.17 fm${}^{-3}$ the intranucleon average 
distance $a_{N}\simeq \rho_{N}^{-1/3}\simeq$ 1.81 fm turns out to be 
remarkably large, a fact that can be appreciated by computing the average 
of $\exp{-{\vert \vec{x}_{1}-\vec{x}_{2}\vert \over{R_{2\pi}}}}$ for 
two nucleons with a gaussian density distribution of radius $R_{N}$ 
which yields the small value 0.05, for $R_{N}\simeq$ 0.7 fm, the nucleon's 
 radius.}. Then what kind of equilibrium 
configuration can we expect for such system ? \par
A dense plasma with a stable neutralizing background bears a close 
resemblance to our nucleon system: in fact the short 
range pionic interaction is a rather accurate mock-up 
of the neutralizing background's interaction with the plasma which is 
Debye-screened at a distance comparable with the distances between 
charges. And the physics of such dense plasma is well known, 
resembling a kind of jelly where the charges, the seeds, oscillate 
around their equilibrium position with the ``plasma 
frequency''\footnote{In order to have some idea about the structure 
of such jelly, it is amusing to pursue in a crude way the analogy with a dense 
plasma. The plasma frequency is then \cite{Preparata2}
\begin{equation}
	\omega_{p} \simeq {g\over {m_{N}^{1/2}}} \left({N\over 
	V}\right)^{1/2} = {g\over {m_{N}^{1/2}}} \left({1\over 
	a_{N}}\right)^{3/2}
\end{equation}
where $g \simeq$ 1 yields at r = $R_{2\pi}$ the reasonable potential 
$V_{2\pi} \simeq$ 100 MeV and $\omega_{p} \simeq$ 40 MeV. The typical 
oscillation amplitude is then $\delta={1\over {(m_{N}\omega_{p})^{1/2}}} 
\simeq 10^{-13}$ cm, a rather reasonable value, too.
}. Wouldn't it,then, be reasonable to expect such a ``jellium'' 
structure to accurately represent the dynamics of the Nucleus as 
well ? To such question the answer of the SM is a surprising, 
incontrovertible no. The nucleons of the Nucleus revolve around it in 
global orbits, much in the same way as the 
electrons whirl around the Nucleus in the Atom. Very strange, isn't 
it ?\par
Indeed, something that would have advised the nuclear physicists to 
look somewhere else in search of a physically realistic basis for the 
remarkable phenomenological success of the SM. From a historical point 
of view it is interesting to contemplate the initial skepticism and 
the later wonder of the leading nuclear 
physicists \cite{Bethe} when confronted with the simplicity and the effectiveness 
of the SM, skepticism and wonder that through habit the successive 
generations came to completely forget, 
deeply involved on one hand in the complicated calculations of 
nuclear structure, and on the other to check the ``self-consistency'' 
of the SM: a task that completely overlooked the fundamental question: 
why a dense plasma, governed in QED by a similar set of interactions 
has a dynamical behaviour which is completely different from that of 
the Nucleus ?\par
The 1990 paper, referred to above, finally succeeded in identifying a 
completely new interaction mechanism, whose basic structure was just 
what is needed to make sense, in a realistic way, of the SM and of 
other suitable aspects of nuclear dynamics\footnote{In \cite{pions} and 
in the Chapter 11 of the book \cite{Preparata2} one can find a number 
of applications of this novel approach}. In a nutshell the fundamental 
idea is that the Nucleon is just one level of the s-wave three 
non-strange quark system, whose excited state is $\Delta$(1232), lying some 
300 MeV above it. These two levels are strongly coupled to the 
$\pi$-field (itself a quark-antiquark system in s-wave), which 
induces the transitions $\pi + N \rightarrow \Delta(1232) \rightarrow 
\pi + N$ etc. The similarity of this physical system with the familiar 
Laser should not escape the attention of anybody. However, in the 
generally accepted view a system of such kind will ``lase'' if and 
only if it is ``inverted'', i.e. if through some suitable device - 
the pump - one brings a large number of atoms to the excited level. 
Furthermore it is important to place the system in a well-tuned 
optical cavity in order to prevent photons to leak out and be lost for 
the coherent laser evolution. If this were always true (it is 
certainly true in the operational conditions of the Lasers) the 
mechanism we are envisaging would be totally 
irrelevant, but it turns out that, contrary to what is generally 
believed, this is not always true. As demonstrated in 1973 by K. Hepp 
and E. Lieb \cite{Hepp}, a system governed by the Dicke Hamiltonian 
\cite{Dicke} (such as the laser) above a certain density and below a 
certain temperature undergoes \textit{spontaneously} a Superradiant Phase 
Transition (SPT) to a Laser-like state, where matter and a number of 
resonant modes of the e.m. field interact coherently, oscillating in 
phase. And this without any need neither of pumps nor of cavities. 
This crucial and revolutionary result, which for mysterious reasons 
has had no impact on the Physics community, was rediscovered and 
generalized by one of us (G.P.) in 1987, and is the focal point of the 
book in Ref.\cite{Preparata2}. Based upon it the Nucleus becomes a 
``\textit{bona-fide}'' Pionic Laser, whose two levels are just the 
Nucleon and the $\Delta$(1232), and, as shown in 
Ref.\cite{Preparata1}, the couplings of both N(940) and 
$\Delta$(1232) to the $\pi$-field are strong enough to 
meet the conditions for a SPT. In this way, 
through the coherent interaction with the $\pi$-field, which gets 
trapped in the region where the collective''N-$\Delta$ current'' is 
localized, i.e. within the Nucleus, the assembly of Nucleons reaches a 
completely novel ground state, where N's and $\Delta$'s oscillate in 
phase and their ``orbit'' are not constrained to be localized, for the 
binding $\pi$-field is spread out throughout the Nucleus, and not 
peaked around the single Nucleons as envisaged by the short-range 
``nucleostatic'' potential. As a matter of fact, as argued in 
Ref.\cite{Preparata1}, the SM just describes the ground state of a 
finite number of Fermions confined by their collective interactions 
within the nuclear volume. In a certain sense we may say that in the new 
approach the Nucleus owes its existence to a ``condensation'' of the 
$\pi$-field within the spatial extent of the Nucleus. But it is clear 
that such ``condensate'' is of a very peculiar type, characterized by 
its well defined phase relation with the N-$\Delta$ oscillations, and 
by the collective, coherent character of its interaction with the 
N-$\Delta$ system.\par
In spite of the remarkably successful phenomenology \cite{pions} 
that one can deduce from the precise quantum field theoretical 
formulation that has been expounded in 
Refs.\cite{Preparata1,Preparata2}, these ideas have found no 
interest nor resonance in the community of nuclear physics. The 
``mystery'' of such a consistent neglect of both the conceptual 
difficulties of nucleostatic forces \textit{vis-\`{a}-vis} 
the SM and the nuclear structure in general, and the satisfactory and 
theoretically compelling solution by the coherent nuclear interaction 
sketched above, has recently been lifted in the occasion of a review 
of our work demanded by a funding Agency. We have finally 
learnt that this approach has been 
foresaken by the community for it violates a well 
know result in Nuclear Physics, which goes back to the beginning of 
the 70's, and is associated mainly with the work of the Russian 
physicist A. B. Migdal \cite{Migdal}. According to this work, which has 
been subsequently refined in many ways\footnote{For a simple but very 
clear account see the book by Ericsson and Weise \cite{Weise}}, 
$\pi$-condensation at the actual nuclear densities 
$\rho_{N}\simeq$ 0.17 fm${}^{-3}$ is ruled out by a strong repulsion 
effects which push the critical density $\rho_{c}\simeq 3\rho_{N}$, 
way above what can be realized in a Nucleus.\par
It is the purpose of the last observation to clarify why the above 
argument is totally irrelevant for assessing the 
validity of the approach of the Coherent Nucleus. In simple terms, as 
described in Ref.\cite{Weise}, the problem of $\pi$-condensation 
is dealt with by analysing the propagator of a $\pi$-field 
$D(\omega, \vec{k}\,)$ in a gas of Nucleons of density $\rho$. The 
condition of condensation is then reduced to finding whether the 
inverse propagator ($\Pi$ is the self energy function)
\begin{equation}
	D^{-1}(\omega, \vec{k}\,)=\omega^2-\vec{k}^{\,2}-m_{\pi}^2-
	\Pi(\omega, \vec{k}\,)
	\label{eq:1}
\end{equation}
has a zero for $\omega \leq$ 0, identifying the critical density 
$\rho_{crit}$ as that density for which the pole of $D(\omega, \vec{k}\,)$ 
is at $\omega$ = 0, i.e.
\begin{equation}
	D^{-1}(0, \vec{k}\,)=-\vec{k}^{\,2}-m_{\pi}^2-\Pi(0, \vec{k}\,) = 0.
	\label{eq:2}
\end{equation}¥
The negative, generally accepted, conclusion about $\pi$-condensation 
in ordinary Nuclei stems from the results of a calculation of the 
$\pi$-propagator which sums \textit{incoherently} the contributions 
(particle-hole) of each of the Nucleons. In this way one obtain for 
the $\pi$-self energy
\begin{equation}
	\Pi(\omega, \vec{k}\,)=-{\vec{k}^{\,2}\chi_{0}(\omega, \vec{k}\,)
	\over {1+g^{\prime}\chi_{0}(\omega, \vec{k}\,)}}
	\label{eq:3}
\end{equation}¥
where the susceptibility function $\chi_{0}(\omega, \vec{k}\,)$ 
receives contributions from both N(940) and $\Delta$(1232) and is 
proportional to the nuclear density $\rho$, as implied by the 
\textit{incoherent} sum. $g^{\prime}$ is the ``correlation 
parameter'', originating from short-range repulsion.\par
It should be now abundantly clear that the pion condensation that is 
predicted by the coherent $\pi$-N-$\Delta$ interaction is totally 
unrelated to the one familiar to the nuclear physicists, and that the 
impossibility of the latter cannot have any bearing 
on the likelihood of the former, which, besides its conceptual 
advantages,has on its side an impressive number of successes 
\cite{Preparata1,Preparata2,pions}.\par
To conclude, whereas \textit{incoherent} $\pi$-condensation is 
definitely ruled out by both theory and experiment, the coherent 
``superradiant'' process that produces a \textit{coherent} 
$\pi$-condensate appears not only solidly rooted 
in theory but also supported by experiments, beginning with the 
stunning effectiveness of the SM.


\begin{references}
\bibitem{Preparata1}
G. Preparata, {\it Il Nuovo Cimento} {\bf 103A} (1990) 1213.
\bibitem{Preparata2}
G. Preparata, {\it QED Coherence in Matter}, World Scientific, 
Singapore (1995).
\bibitem{Maier}
M. G. Mayer, {\it Phys. Rev.} {\bf 75} (1949) 1969;\hfill\break
O. Haxel, J. H. Jensen and H: E. Suess, {\it Phys. Rev.} {\bf 75} 
(1949) 1766.
\bibitem{Yukawa}
H. Yukawa, {\it Proc. Phys. Math. Soc. of Japan} {\bf 17} (1935) 48
\bibitem{Machleidt}
R. Machleidt, {\it Adv. Nucl. Phys.} {\bf 19} (1989) 189.
\bibitem{Bethe}
H. A. Bethe, {\it Sci. Am.} 189 (1953) 58;\hfill\break
{\it Phys. Rev.} {\bf 103} (1956) 1352.
\bibitem{pions}
G. Preparata and P. G. Ratcliffe, {\it Il Nuovo Cimento} {\bf A106} 
(1993) 685;\hfill\break
R. Alzetta, M. Gibilisco, G. Liberti and G. Preparata, {\it Mod. Phys. 
Lett.} {\bf A8} (1993) 2335;\hfill\break
E. Del Giudice, R. Mele, G. Preparata, C. Gualdi, G. Mangano and G. 
Miele, {\it Int. Jou. Mod. Phys.} {\bf D4} (1995) 531;\hfill\break
R. Alzetta, G. Liberti and G. Preparata {\it Nucl. Phys.} {\bf A585} 
(1995) 307c;\hfill\break
R. Alzetta, T. Bubba, G. Liberti and G. Preparata {\it Il Nuovo Cimento} 
{\bf 110A} (1997) 169;\hfill\break
R. Alzetta, R. Le Pera, G. Liberti and G. Preparata {\it Il Nuovo Cimento} 
{\bf 110A} (1997) 179;\hfill\break
R. Alzetta, T. Bubba, R. Le Pera, G. Liberti, G. Mileto, D. Tarantino 
and G. Preparata {\it Coherent QED, Giant Resonances and 
(e${}^{+}$e${}^{-}$) Pairs in High Energy Nucleus-Nucleus 
Colllisions}, preprint MITH98/9, NUCL-TH/9811020 (submitted to {\it 
Phys. Rev.} C).
\bibitem{Hepp}
K. Hepp and E. Lieb, {\it Ann. of Phys.} {\bf 76} (1973) 360; {\it 
Phys. Rev.} {\bf A8} (1973) 2517.
\bibitem{Dicke}
R. H. Dicke, {\it Phys. Rev.} {\bf 89} (1953) 472.
\bibitem{Migdal}
A. B. Migdal, {\it Sov. Phys. JEPT} {\bf 36} (1973) 1052;\hfill\break
{\it Revs. Mod. Phys.} {\bf 50} (1978) 10.
\bibitem{Weise}
T. Ericsson and W. Weise, {\it Pions and Nuclei},Oxford 
(1988);\hfill\break
M. Rho and D. H. Wilkinson (Eds.), {\it Mesons in Nuclei}, Amsterdam 
(1979).

\end{references}
\end{document}